\documentclass[12pt, a4paper]{article}
\pdfoutput=1
\usepackage{graphicx}
\usepackage{amssymb}
\usepackage{amsmath}
\usepackage{bm}
\usepackage[table,dvipsnames]{xcolor} %for \rowcolors
\usepackage{cite}
\usepackage{slashed}
\usepackage{epstopdf}            
\usepackage{epsfig}
\usepackage{here}
\usepackage{comment}
\usepackage{booktabs} %for \toprule
\usepackage{colortbl} %for \rowcolor
\usepackage{wrapfig}
\usepackage{ascmac}
\usepackage{fancybox}  
\usepackage{soul} % strike through text
\usepackage{adjustbox}

\allowdisplaybreaks
\renewcommand{\thefootnote}{\fnsymbol{footnote}}
\numberwithin{equation}{section} % Eq.(Sec.eq.)
\def\beq#1\eeq{\begin{align}#1\end{align}}
\newcommand{\ov}{\overline}

\renewcommand{\arraystretch}{1.3}
\RequirePackage{xspace}
\usepackage{caption}
\definecolor{BlueViolet}{rgb}{0.2, 0.00, 0.7}
\definecolor{Blue}{rgb}{0.15, 0.00, 0.9}
\definecolor{light_blue}{rgb}{0.15, 0.35, 0.95}
\definecolor{kit_green}{rgb}{0
, 0.58823 %150/255
, 0.50980 %130/255
}
\usepackage[%dvipdfmx,
colorlinks=true, linkcolor=light_blue,citecolor=light_blue,urlcolor=kit_green]{hyperref} 
\usepackage[normalem]{ulem}
\begin{document}
\sloppy %https://tex.stackexchange.com/questions/9107/how-can-i-make-my-text-never-go-over-the-right-margin-by-always-hyphenating-or-b
\begin{titlepage}
\begin{center}
%%%%%%%%%%%%%%%%%%%%%%%%%%%%%%%%%%%%%%%%%%%
\hfill{KEK--TH--2818, P3H--26--016, TTP26--005}\\
\vskip .3in

{\Large{\bf $b \to c$ semileptonic
sum rule:\\ \vskip .04in
exploring a sterile neutrino loophole\\ \vskip .04in
}}

\vskip .3in

% bold applies to math too
\makeatletter\g@addto@macro\bfseries{\boldmath}\makeatother

{ \large
Motoi Endo$^{\rm a,b,c}$,\,
Syuhei Iguro$^{\rm d,c}$,\,
Tim Kretz$^{\rm e}$,\,
Satoshi Mishima$^{\rm f}$
}
\vskip .3in
$^{\rm a}${\it KEK Theory Center, IPNS, KEK, Tsukuba 305--0801, Japan}\\\vspace{4pt}
$^{\rm b}${Graduate Institute for Advanced Studies, SOKENDAI, Tsukuba,\\ Ibaraki 305--0801, Japan} \\\vspace{4pt}
$^{\rm c}${\it Kobayashi-Maskawa Institute (KMI) for the Origin of Particles and the Universe, Nagoya University, Nagoya 464--8602, Japan}\\\vspace{4pt}
$^{\rm d}${\it Institute for Advanced Research (IAR), Nagoya University,\\ Nagoya 464--8601, Japan}\\\vspace{4pt}
$^{\rm e}${\it Institute for Theoretical Particle Physics (TTP), Karlsruhe Institute of Technology (KIT), Wolfgang-Gaede-Str.\,1, 76131 Karlsruhe, Germany}\\\vspace{4pt}
$^{\rm f}${\it Department of Liberal Arts, Saitama Medical University, Moroyama,\\ Saitama 350-0495, Japan}
\end{center}
\vskip .15in

%%%%%%%%%%%%%%%%%%%%%%%%%
\begin{abstract}
We investigate the $b\to c$ semileptonic sum rule in the presence of a massive sterile neutrino. 
Recent measurements of charged-current semitauonic $B$-meson decays exhibit a $\sim4\sigma$ deviation from the Standard Model predictions, whereas no such tension has been reported for the lowest-lying baryonic counterpart, the $\Lambda_b$ decay. 
Since these decay rates are related through the sum rule, accommodating such a mismatch beyond the level of uncertainties is nontrivial.
We revisit this issue by considering dimension-six operators involving a massive sterile neutrino, and evaluate the resulting violation of the sum rule.
We find that the induced effect remains negligible compared with the current experimental uncertainties, further strengthening the sum rule as a consistency check for the experimental data.
%
%%%%%%%%%%%%%%%%%%%%%%%%%
\end{abstract}
\end{titlepage}

\setcounter{page}{1}
\renewcommand{\thefootnote}{\#\arabic{footnote}}
\setcounter{footnote}{0}

%%%%%%%%%%%%%%%%%%%%%%%%%
% Contents
%%%%%%%%%%%%%%%%%%%%%%%%%
\hrule
\tableofcontents
\vskip .2in
\hrule
\vskip .4in
%%%%%%%%%%%%%%%%%%%%%%%%%

%%%%%%%%%%%%%%%%%%%%%%%%%%%%%%%
\section{Introduction}
\label{sec:intro}
%%%%%%%%%%%%%%%%%%%%%%%%%%%%%%%
Lepton-flavor universality is a fundamental prediction of the Standard Model (SM).
The semitauonic $B$-meson decays, $B \to D^{(*)} \tau \ov{\nu}$, have attracted attention since the measurements of $R_{D^{(*)}} ={\rm{BR}}(B\to D^{(*)} \tau\bar\nu_\tau)/{\rm{BR}}(B\to D^{(*)} \ell\bar\nu_\ell)$, with $\ell=e,\mu$, have shown deviations of about $4\sigma$ from the SM predictions \cite{HFLAV:2024ctg}.
These deviations may hint at the presence of new physics (NP) \cite{Iguro:2024hyk}, and the Belle II and LHCb experiments are accumulating data to further scrutinize this possibility \cite{ATLAS:2025lrr}.
In contrast, a recent measurement of the baryonic counterpart, $R_{\Lambda_c} ={\rm{BR}}(\Lambda_b\to\Lambda_c\tau\ov\nu_\tau)/{\rm{BR}}(\Lambda_b\to\Lambda_c\ell\ov\nu_\ell)$, reported by the LHCb collaboration~\cite{LHCb:2022piu} is consistent with the SM prediction within uncertainties.

A robust cross-check of the experimental results is provided by the $b \to c$ semileptonic sum rules, which relate the decay rates of $\Lambda_b \to \Lambda_c \tau\bar\nu$ and $B \to D^{(*)} \tau\bar\nu$ \cite{Blanke:2018yud,Blanke:2019qrx,Fedele:2022iib,Duan:2024ayo}, as
\begin{align}
 \frac{R_{\Lambda_c}}{R_{\Lambda_c}^{\rm SM}} - \alpha \frac{R_{D}}{R_{D}^{\rm SM}} - \beta \frac{R_{D^*}}{R_{D^*}^{\rm SM}}=\delta,
 \label{eq:RSR}
\end{align}
where the coefficients satisfy $\alpha + \beta = 1$.\footnote{
A similar sum rule also holds for $B_s\to D_s^{(*)} \tau\ov\nu$ and $\Xi_b\to\Xi_c \tau\ov\nu$, which are related by SU(3) flavor symmetry to $B \to D^{(*)} \tau\bar\nu$ and $\Lambda_b \to \Lambda_c \tau\bar\nu$, respectively \cite{Iguro:2026xgi}.}
In the small-velocity (Shifman-Voloshin) limit, where the heavy-quark and zero-recoil limits are taken, one obtains $\alpha = 1/4$ and $\beta = 3/4$ with $\delta = 0$ \cite{Endo:2025fke, Endo:2025lvy}.
For realistic hadrons, however, the limit is violated, and $\delta$ parametrizes the violation of the sum rule.
Such a violation generally requires NP interactions and arises through the masses of the beauty and charming hadrons (denoted as $H_b$ and $H_c$ respectively) as well as via the $H_b\to H_c$ transition form factors \cite{Endo:2025fke}.
It has been shown that the magnitude of $\delta$ is negligible compared with the current experimental uncertainties \cite{Endo:2025lvy}.
On the other hand, if we substitute the current experimental values of $R_{\Lambda_c}$ \cite{LHCb:2022piu}, $R_{D}$ and $R_{D^*}$ \cite{HFLAV:2024ctg} into the left-hand side of Eq.~\eqref{eq:RSR}, the parameter is estimated as $\delta\simeq-0.41\pm 0.24$.\footnote{
This value mildly depends on the form factor parameterization. 
See Ref.~\cite{Endo:2025lvy} for details.}
This result implies a mild tension among the experimental results, necessitating further investigation.

The above conclusion has been drawn under the assumption that the neutrino involved is always left-handed and massless.
However, since neutrinos are not detected directly in the above measurements, their chirality and mass have not been identified.
In this work, we introduce a massive right-handed neutrino ($N_R$, hereafter referred to as a sterile neutrino).
In addition to the SM contribution, a sterile neutrino can be emitted in the semileptonic decay, and the total decay rate is written as
\begin{align}
 \Gamma^\prime_{H_c} \equiv \Gamma(H_b\to H_c \tau \bar\nu_\tau)+\Gamma(H_b\to H_c \tau \bar N_R).
 \label{eq:Gamma}
\end{align}
As argued in Refs.~\cite{Endo:2025set, Endo:2025lvy}, the sum rule is expected to hold exactly in the small-velocity limit even with a sterile neutrino.
For realistic hadrons, however, the relation is violated, and a massive $N_R$ could affect the size of this violation in two ways: (i) the sterile neutrino mass enters the differential decay rates, and (ii) a sufficiently heavy $N_R$ can kinematically close individual decay channels, invalidating the sum rule construction entirely.
Motivated by these considerations, we test whether a massive sterile neutrino can provide a viable loophole to the $b \to c$ semileptonic sum rule.

The structure of the paper is given as follows.
In Sec.~\ref{sec:model}, we outline the effective operators describing NP contributions involving a sterile neutrino. 
Section~\ref{sec:Formulation} describes the decay rate and Sec.~\ref{sec:Numerical} discusses the sum rule violation as well as its implications.
Section~\ref{sec:conclusion} is devoted to concluding our findings and discussion.
%

%%%%%%%%%%%%%%%%%%%%%%%%%%%%%%%
\section{New physics framework}
\label{sec:model}
%%%%%%%%%%%%%%%%%%%%%%%%%%%%%%%

In this paper, we assume that NP contributions appear as $b\to c \tau\bar N_R$, where $N_R$ is a massive right-handed neutrino. 
The effective Hamiltonian is introduced as 
\begin{align}
 \label{eq:Hamiltonian}
 {\cal {H}}_{\rm{eff}}= 2 \sqrt2 \, G_F V_{cb}\biggl[ O_{V_L}+C_{V_R^\prime} O_{V_R^\prime}+C_{S_L^\prime} O_{S_L^\prime}+C_{S_
R^\prime} O_{S_R^\prime}+C_{T^\prime} O_{T^\prime} \biggr].
\end{align}
Here, $O_{V_L} = (\overline{c}\gamma^\mu P_L b)(\overline{\tau}\gamma_\mu P_L \nu_\tau)$ 
denotes the SM contribution. 
The dimension-six effective operators involving $N_R$ are given by
\begin{align}
 &O_{V_R^\prime} = (\overline{c} \gamma^\mu P_R b)(\overline{\tau} \gamma_\mu P_R N_R),
 &&O_{S_R^\prime} = (\overline{c}  P_R b)(\overline{\tau} P_R N_R), \label{eq:operator1}\notag \\
 &O_{S_L^\prime} = (\overline{c} P_L b)(\overline{\tau} P_R N_R),
 &&O_{T^\prime} = (\overline{c}  \sigma^{\mu\nu}P_R b)(\overline{\tau} \sigma_{\mu\nu} P_R N_R),
\end{align}
where $P_{L(R)}=(1\mp\gamma_5)/2$ are the chirality projection operators.
Such contributions have been discussed, for instance, in Refs.~\cite{Iguro:2018qzf,Robinson:2018gza,Babu:2018vrl, Mandal:2020htr,Penalva:2021wye,Datta:2022czw}. 
The NP contributions are encoded in the Wilson coefficients (WCs) $C_{X^\prime}$, which are normalized with the factor $2 \sqrt2 G_F V_{cb}$.
In contrast to the conventional scenarios, we focus on NP contributions arising from the sterile neutrinos.
In this framework, the SM limit corresponds to $C_{X^\prime} = 0$ for $X=V_{R}$, $S_{L,R}$, and $T$.
In the following, we consider decays into a massive sterile neutrino and assume that the emitted $N_R$ does not decay inside the detector, thereby contributing to missing energy.

%%%%%%%%%%%%%%%%%%%%%%%%%%%%%%%
\section{Sum rule with a sterile neutrino}
\label{sec:Formulation}
%%%%%%%%%%%%%%%%%%%%%%%%%%%%%%%

Schematically, the decay amplitude can be written in the following form,
\begin{align}
\mathcal{M}=\frac{G_F V_{cb}}{\sqrt{2}} \,H \times L,
\end{align}
where $H$ and $L$ denote the hadronic and leptonic amplitudes, respectively.
For simplicity, the polarization vectors, the summation over polarizations, the polarization labels of the daughter hadron and charged lepton, as well as Lorentz indices are omitted, see Ref.~\cite{Tanaka:2012nw}.
In contrast to the decay into a left-handed neutrino, the difference arises in the leptonic amplitude.
The explicit expressions involving a sterile neutrino are given in appendix \ref{sec:amp}. 

In terms of the effective Hamiltonian of Eq.~(\ref{eq:Hamiltonian}), we provide the differential decay rates as \cite{Gratrex:2015hna}
\begin{align}
 &\frac{d\Gamma (B \to D \tau\bar N_R)}{dq^2} =N_b \biggl{\{}
 3|C_{S_R^\prime}+C_{S_L^\prime}|^2 \left(q^2-m_\tau^2-m_{N_R}^2 \right) (H^{D}_S)^2 
 \label{eq:BD} \\ \notag
 &~~~~~~~+16|C_{T^\prime}|^2 \frac{\left( q^2\right)^2 +q^2(m_\tau^2+m_{N_R}^2) -2(m_\tau^2-m_{N_R}^2)^2}{q^2}(H^{D}_{T})^2 \\ \notag
 &~~~~~~~+|C_{V_R^\prime}|^2 \biggl[\frac{3 \left(q^2(m_\tau^2+m_{N_R}^2)-(m_\tau^2-m_{N_R}^2)^2\right)}{q^2} (H^{D}_{V,t})^2\\ \notag
 &~~~~~~~~~~~~~~~~~~~~ +\frac{ 2\left( q^2\right)^2 -q^2(m_\tau^2+m_{N_R}^2)-(m_\tau^2-m_{N_R}^2)^2  }{q^2} (H^{D}_{V,0})^2 \biggr{]} \\ \notag
 &~~~~~~~+6{\rm{Re}} \biggl{[} (C_{S_R^\prime}+C_{S_L^\prime}) C_{V_R^\prime}^* \biggr{]} \frac{m_\tau (q^2-m_\tau^2+m_{N_R}^2)}{\sqrt{q^2}} H^{D}_S H^{D}_{V,t}\\ \notag
 &~~~~~~~-24{\rm{Re}} \biggl{[} C_{T^\prime} C_{V_R^\prime}^{*} \biggr{]} \frac{m_\tau(q^2-m_\tau^2+m_{N_R}^2)}{\sqrt{q^2}}H^{D}_{T}H^{D}_{V,0}\,\biggr{\}},\notag
\end{align}
%%%%%

%%%%%
\begin{align}
 &\frac{d\Gamma (B \to D^* \tau\bar N_R)}{dq^2} =N_b \biggl{\{}
 3|C_{S_R^\prime}-C_{S_L^\prime}|^2 \left(q^2-m_\tau^2-m_{N_R}^2 \right) (H^{D^*}_P)^2~~~
 \label{eq:BDs} \\ \notag
 &~~~~~~~+16|C_{T^\prime}|^2 \frac{\left( q^2\right)^2 +q^2(m_\tau^2+m_{N_R}^2) -2(m_\tau^2-m_{N_R}^2)^2}{q^2} \left((H^{D^*}_{T,0})^2+(H^{D^*}_{T,+})^2+(H^{D^*}_{T,-})^2 \right) \\ \notag
 &~~~~~~~+|C_{V_R^\prime}|^2 \biggl[\frac{3\left(q^2(m_\tau^2+m_{N_R}^2)-(m_\tau^2-m_{N_R}^2)^2\right)}{q^2} (H^{D^*}_{V,t})^2\\ \notag
 &~~~~~~~~~~~~~~~~~ +\frac{ 2\left( q^2\right)^2 -q^2(m_\tau^2+m_{N_R}^2)-(m_\tau^2-m_{N_R}^2)^2  }{q^2} \left((H^{D^*}_{V,0})^2+(H^{D^*}_{V,+})^2+(H^{D^*}_{V,-})^2 \right) \biggr{]} \\ \notag
  &~~~~~~~-6{\rm{Re}} \biggl{[} (C_{S_R^\prime}-C_{S_L^\prime}) C_{V_R^\prime}^* \biggr{]} \frac{ m_\tau (q^2-m_\tau^2+m_{N_R}^2)}{\sqrt{q^2}} H^{D^*}_P H^{D^*}_{V,t}\\ \notag
  &~~~~~~~-24{\rm{Re}} \biggl{[} C_{T^\prime} C_{V_R^\prime}^* \biggr{]} \frac{m_\tau(q^2-m_\tau^2+m_{N_R}^2)}{\sqrt{q^2}}(H^{D^*}_{T,0}H^{D^*}_{V,0}+H^{D^*}_{T,+}H^{D^*}_{V,+}-H^{D^*}_{T,-}H^{D^*}_{V,-})\,\biggr{\}}, \notag
\end{align}
%%%%%

%%%%%
\begin{align}
 &\frac{d\Gamma ( \Lambda_b \to \Lambda_c \tau\bar N_R)}{dq^2} =\frac{N_b}{2} \biggl{\{} 6|C_{S_R^\prime}+C_{S_L^\prime}|^2(q^2-m_\tau^2-m_{N_R}^2)(H^{\Lambda_c}_{S})^2
  \label{eq:Lambdabc} \\ \notag
 &~~~~~~~~~+6|C_{S_R^\prime}-C_{S_L^\prime}|^2(q^2-m_\tau^2-m_{N_R}^2)(H^{\Lambda_c}_{P})^2 \\ \notag
 &~~~~~~~~~+16|C_{T^\prime}|^2  \frac{\left(q^2\right)^2+q^2(m_\tau^2+m_{N_R}^2)-2(m_\tau^2-m_{N_R}^2)^2}{q^2} \\ \notag
 &~~~~~~~~~~~~~~~~~~~~~~~~~~~~~~~~~~~~~\left((H^{\Lambda_c}_{T,0+})^2+(H^{\Lambda_c}_{T,0-})^2+(H^{\Lambda_c}_{T,1+})^2+(H^{\Lambda_c}_{T,1-})^2\right) \\ \notag
 &~~~~~~~~~+|C_{V_R^\prime}|^2 \biggl{[} \frac{ 2\left( q^2\right)^2 -q^2(m_\tau^2+m_{N_R}^2)-(m_\tau^2-m_{N_R}^2)^2  }{q^2}\\ \notag
 &~~~~~~~~~~~~~~~~~~~~~~~~~~~~~~~~~~~~~\left((H^{\Lambda_c}_{V,0+})^2+(H^{\Lambda_c}_{V,0-})^2+(H^{\Lambda_c}_{V,1+})^2+(H^{\Lambda_c}_{V,1-})^2 \right) \\ \notag
 &~~~~~~~~~~~~~~~~~~~~+3\frac{q^2(m_\tau^2+m_{N_R}^2)-(m_\tau^2-m_{N_R}^2)^2}{q^2}\left((H^{\Lambda_c}_{V,t+})^2+(H^{\Lambda_c}_{V,t-})^2 \right) \biggr{]}\\ \notag
 &~~~~~~~~~+6{\rm{Re}} \biggl{[} (C_{S_R^\prime}+C_{S_L^\prime}) C_{V_R^\prime}^* \biggr{]}\frac{m_\tau(q^2-m_\tau^2+m_{N_R}^2)}{\sqrt{q^2}} H^{\Lambda_c}_{S}\left( H^{\Lambda_c}_{V,t+}+H^{\Lambda_c}_{V,t-}\right) \\ \notag
 &~~~~~~~~~+6{\rm{Re}} \biggl{[} (C_{S_R^\prime}-C_{S_L^\prime}) C_{V_R^\prime}^* \biggr{]}\frac{m_\tau(q^2-m_\tau^2+m_{N_R}^2)}{\sqrt{q^2}} H^{\Lambda_c}_{P}\left( H^{\Lambda_c}_{V,t+}-H^{\Lambda_c}_{V,t-}\right) \\ \notag
 &~~~~~~~~~+24 {\rm{Re}}\biggl{[} C_{T^\prime} C_{V_R^\prime}^* \biggr{]} \frac{m_\tau(q^2-m_\tau^2+m_{N_R}^2)}{\sqrt{q^2}}  \\ \notag
 &~~~~~~~~~~~~~~~~~~~~~~~~~~~~~~~~(H^{\Lambda_c}_{T,0+}H^{\Lambda_c}_{V,0+}+H^{\Lambda_c}_{T,0-}H^{\Lambda_c}_{V,0-}+H^{\Lambda_c}_{T,1+}H^{\Lambda_c}_{V,1+}+H^{\Lambda_c}_{T,1-}H^{\Lambda_c}_{V,1-})\, 
 \biggr{\}}, \notag
\end{align}
where $N_b$, $Q_{\pm}^q$, and $Q_{\pm}^l$ are defined as
\begin{align}
N_b=&\frac{G_F^2|V_{cb}|^2 %\eta_{EW}^2 
\sqrt{Q_{+}^qQ_{-}^qQ_{+}^lQ_{-}^l}}{384\pi^3 m_{H_b}^3q^2},\\
Q_{\pm}^q=(m_{H_b}\pm &m_{H_c})^2-q^2,\,\,\,Q_{\pm}^l=(m_{\tau}\pm m_{N_R})^2-q^2.
\end{align}
In addition, the squared invariant mass of the lepton system is defined as $q^2=(p_\tau+p_{N_R})^2$.

The above expressions are valid for realistic hadrons, and are significantly simplified in the heavy quark limit.
In the HQET, the hadron masses are expressed as \cite{Falk:1992wt,Falk:1992ws}
\begin{align}
 m_{H_Q} = m_Q + \bar \Lambda + \frac{\Delta m^2}{2m_Q} + \ldots \,,
 \label{eq:mass_HQET}
\end{align}
where $m_Q$ denotes the heavy quark mass, and $\bar \Lambda$ and $\Delta m^2$ are QCD-scale parameters. 
In the heavy quark limit, $m_Q \gg \bar \Lambda$, so that the hadron masses are well approximated by the heavy quark mass.
Moreover, in this limit, the heavy quark symmetry is restored, and the form factors are described by the leading-order Isgur–Wise (IW) functions, while higher-order corrections are suppressed.
Denoting the leading IW functions for the $\Lambda_b\to\Lambda_c$ and $B\to D^{(*)}$ transitions by $\zeta$ and $\xi$, respectively, one obtains the following relation analytically in the heavy quark limit, as anticipated in Ref.~\cite{Endo:2025set},
\begin{align}
  \frac{\kappa_{\Lambda_c}}{\zeta(w)^2}= \frac{2}{w+1} \frac{\kappa_{D}+\kappa_{D^*}}{\xi(w)^2},
  \label{eq:KappaSR}
\end{align}
where $\kappa_{H_c}=d\Gamma(H_b\to H_c\tau \bar N_R)/dq^2$ and 
$w=\left(m_{H_b}^2+m_{H_c}^2-q^2\right)/\left(2m_{H_b}m_{H_c}\right)$.\footnote{
Since $q^2$ is more convenient for discussing the kinematic threshold for $N_R$ emission, $q^2\ge (m_\tau+m_{N_R})^2$, we hereafter use $q^2$ rather than $w$, unless otherwise stated explicitly.
}

Equation \eqref{eq:KappaSR} also holds for decays into a left-handed neutrino instead of a sterile neutrino \cite{Endo:2025fke}.
The sum rule for the differential decay rates then reads
\begin{align}
  \frac{\kappa^\prime_{\Lambda_c}}{\zeta(w)^2}= \frac{2}{w+1} \frac{\kappa^\prime_{D}+\kappa^\prime_{D^*}}{\xi(w)^2},
  \label{eq:KappaSRall}
\end{align}
where $\kappa^\prime_{H_c} = d\Gamma(H_b\to H_c\tau \bar \nu_\tau)/dq^2 + d\Gamma(H_b\to H_c\tau \bar N_R)/dq^2$.
Integrating both sides over the phase space and following the discussion in Ref.~\cite{Endo:2025lvy}, one obtains the relation for the total decay rates at leading order in the small-velocity limit,
\begin{align}
  \Gamma^\prime_{\Lambda_c} = \Gamma^\prime_{D}+\Gamma^\prime_{D^*}.
  \label{eq:GammaSR}
\end{align}
Here, the decay rates include contributions from both left-handed and sterile neutrino emissions (see Eq.~\eqref{eq:Gamma}).
In this limit, the SM contributions satisfy the alternative relations, $\Gamma_{D}^{\rm SM}/\Gamma_{\Lambda_c}^{\rm SM} = 1/4$ and $\Gamma_{D^*}^{\rm SM}/\Gamma_{\Lambda_c}^{\rm SM} = 3/4$ \cite{Endo:2025lvy}.
Using Eq.~\eqref{eq:GammaSR}, one then obtains the following relation for the total decay rates,
\begin{align}
 \frac{\Gamma^\prime_{\Lambda_c}}{\Gamma_{\Lambda_c}^{\rm SM}} = 
 \frac{1}{4} \frac{\Gamma^\prime_{D}}{\Gamma_{D}^{\rm SM}} + \frac{3}{4} \frac{\Gamma^\prime_{D^*}}{\Gamma_{D^*}^{\rm SM}}.
 \label{eq:GammaSR2}
\end{align}
Finally, by normalizing each decay rate with that of the light lepton mode $\Gamma(H_b\to H_c \ell \bar\nu_\ell)$, we obtain the $b \to c$ semileptonic sum rule for decays involving sterile neutrinos,
\begin{align}
 \frac{R^\prime_{\Lambda_c}}{R_{\Lambda_c}^{\rm SM}} = \frac{1}{4} \frac{R^\prime_{D}}{R_{D}^{\rm SM}} + \frac{3}{4} \frac{R^\prime_{D^*}}{R_{D^*}^{\rm SM}}.
 \label{eq:RSR_limit}
\end{align}
The ratios $R_{H_c}^\prime$ include contributions from both left-handed and sterile neutrinos,
\begin{align}
    R_{H_c}^\prime=\frac{{\rm{BR}}(H_b\to H_c \tau \bar\nu_\tau)+{\rm{BR}}(H_b\to H_c \tau \bar N_R)}{{\rm{BR}}(H_b\to H_c \ell\bar\nu_\ell)}.
\end{align}

The equality in Eq.~\eqref{eq:RSR_limit} holds exactly in the small-velocity limit.
For realistic hadrons, however, this limit is violated.
We introduce $\delta^{N_R}$ to parametrize the deviation associated with sterile neutrino contributions,
\begin{align}
 \delta^{N_R} = \frac{R^\prime_{\Lambda_c}}{R_{\Lambda_c}^{\rm SM}} - \frac{1}{4} \frac{R^\prime_{D}}{R_{D}^{\rm SM}} - \frac{3}{4} \frac{R^\prime_{D^*}}{R_{D^*}^{\rm SM}}.
 \label{eq:RSRHQw1}
\end{align}
As discussed in Ref.~\cite{Endo:2025fke}, $\delta^{N_R}$ vanishes in the absence of sterile neutrino contributions, $R_{H_c}^\prime = R_{H_c}^{\rm SM}$.
The quantity $\delta^{N_R}$ can therefore be expanded in terms of the Wilson coefficients describing NP contributions, 
\begin{align}
 \delta^{N_R} \equiv \sum_{ij} \delta_{ij}^{N_R}\,C_i\,C_j^{*}.~~~
 (i,j = V_R^\prime, S_L^\prime, S_R^\prime, T^\prime)
 \label{eq:RSRHQw2}
\end{align}

As mentioned above, the deviation $\delta^{N_R}$ arises from the violation of the small-velocity limit, {\it i.e.}, through corrections associated with the hadron masses and the form factors.
In contrast to the case of left-handed neutrino emission, the sterile neutrino mass appears explicitly in the differential decay rates in Eqs.~\eqref{eq:BD}–\eqref{eq:Lambdabc}, thereby affecting $\delta^{N_R}$.
Moreover, since the hadron masses are no longer approximated by the heavy quark masses and depend on the hadron species, some of the decay channels can become kinematically forbidden if the sterile neutrino is sufficiently heavy.
In particular, the channels close for $m_{N_R}\gtrsim 1.64$, $1.49$, and $1.56\,{\rm GeV}$ for $B\to D\tau\bar N_R$, $B\to D^*\tau\bar N_R$, and $\Lambda_b \to \Lambda_c\tau\bar N_R$, respectively.
Once any of these channels is kinematically closed, the construction of the sum rule no longer applies, and the equality in Eq.~\eqref{eq:RSR_limit} is violated.

%%%%%%%%%%%%%%%%%%%%%%%%%%%%%%%
\section{Numerical result}
\label{sec:Numerical}
%%%%%%%%%%%%%%%%%%%%%%%%%%%%%%%

In this section, we examine the sum rule violation $\delta^{N_R}$ and explore how large it can be.
The hadronic transition form factors are evaluated using the HQET, including higher-order corrections, following Refs.~\cite{Iguro:2020cpg,Bernlochner:2018kxh}.\footnote{
For simplicity, we neglect theoretical uncertainties, in particular those associated with the form factors, since they are expected to be small when using HQET-based form factors (see, {\it e.g.}, Ref.\cite{Endo:2025lvy}). 
A detailed treatment of these uncertainties is left for future work.
}

%%%%%%%%%%%%%%%%%%%%%%%%%%%%
\begin{table}[t]
  \renewcommand{\arraystretch}{1.3}
  \centering
  \begin{tabular}{ccccc}
  & $C_{V_R^\prime}$ & $C_{S_L^\prime}$ & $C_{S_R^\prime}$ & $C_{T^\prime}$  \\    
  \hline
 EFT ($>10\,\text{TeV}$) &  $  0.33 $ &$   0.55 $ & $ 0.55 $ &  $  0.17$ \\ %\hline
LQ ($2\,\text{TeV}$)  & $ 0.51 $ & $ 0.80 $ & $ 0.77 $ & $ 0.30$ \\ 
    \hline
  \end{tabular}
\caption{\label{tab:LHC_bound}
The $95\%$ CL upper bounds on the WCs at the bottom quark scale $\mu = \mu_b$, obtained from the reinterpretation of the high-$p_T$ mono-$\tau$ search.
The NP mass scale is taken as $M_{\rm LQ} = 2$\,TeV with $\Lambda_{\rm EFT} >10\,$TeV.
The bounds are independent of the sign and complex phase of the WCs.
}
\end{table}
%%%%%%%%%%%%%%%%%%%%%%%%%%%%

As shown in Eq.~\eqref{eq:RSRHQw2}, the parameter $\delta_{ij}^{N_R}$ enters $\delta^{N_R}$ weighted by the Wilson coefficients $C_i$.
The bound on $C_i$ is obtained by reinterpreting the LHC constraint on $bc\tau\nu$ four-Fermi operators \cite{Iguro:2020keo}.
The effect of the sterile neutrino mass is negligible, since the mass considered in our analysis is much smaller than the experimental resolution.
The bound is derived assuming the $t$-channel exchanges of leptoquarks (LQs).
Table~\ref{tab:LHC_bound} summarizes the current constraint obtained from the high-$p_T$ mono-$\tau$ search.
It should be noted that the bound depends on the mass scale of the mediating LQ.
In the following, we refer to the case with a 2\,TeV LQ mediator as the relaxed LHC bound, while the EFT-like limit is referred to as the stringent LHC bound.
In general, scenarios with $s$-channel mediators are more strongly constrained due to the peak-like structure in the signal events \cite{Iguro:2018fni}, whereas scenarios with mediators at the electroweak scale are more difficult to constrain because of the large background rates \cite{Iguro:2022uzz,Blanke:2022pjy}.
However, these constraints are highly model dependent, and therefore we adopt the bounds summarized in Table~\ref{tab:LHC_bound}.
Owing to SU(2)$_L$ invariance, these operators also contribute to other observables, which allows us to derive upper limits on the corresponding Wilson coefficients.
We return to this point at the end of this section and impose the LHC bound as a minimal requirement.
%%%%%%%%%%%%%%%%%%%%%%%%%%%%%%%%%%%%%%%
\begin{figure}[t]
\begin{center}
\includegraphics[width=0.315 
\linewidth]{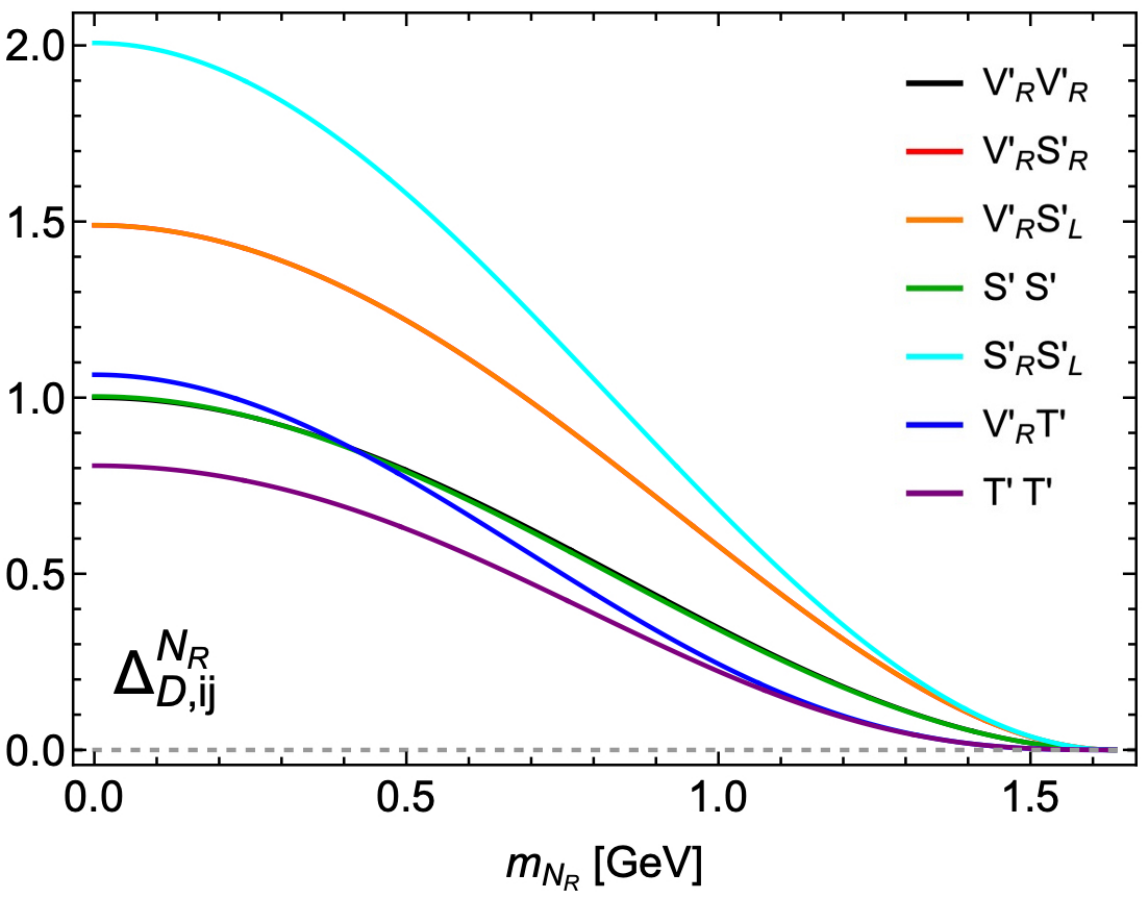}~
\includegraphics[width=0.325  
\linewidth]{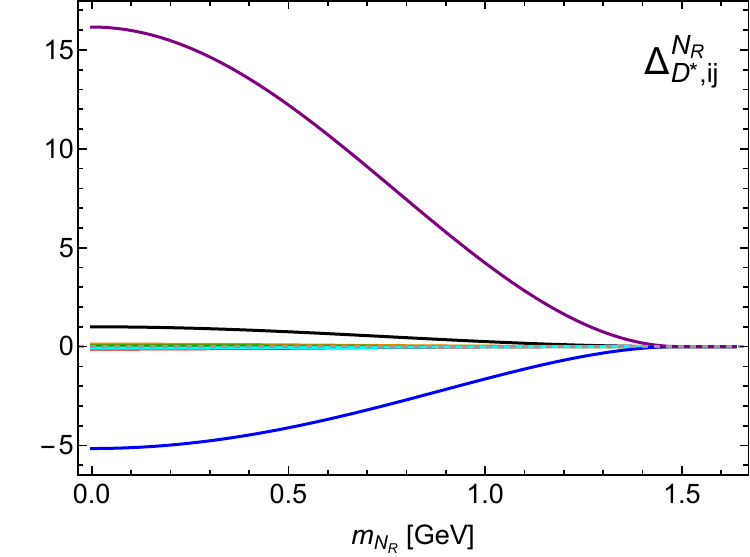}~
\includegraphics[width=0.32\linewidth]{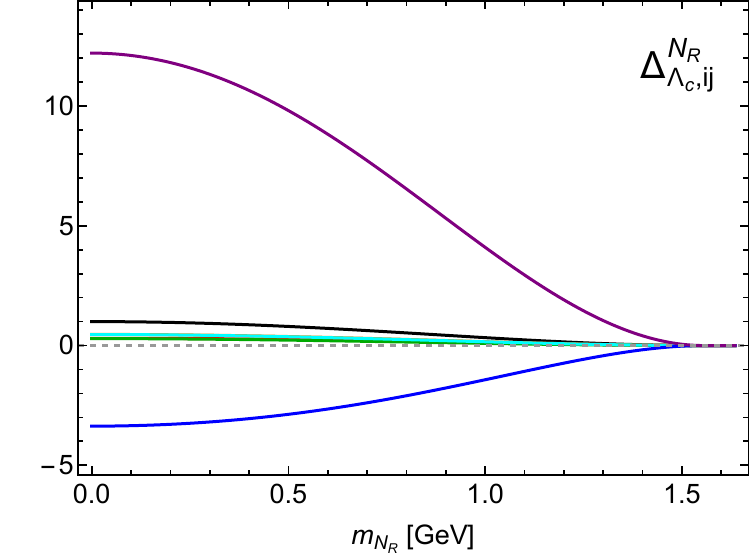}\\
\end{center}
\vspace{-.15cm}
\caption{
The parameters $\Delta_{H_c,\,ij}^{N_R}$ as functions of $m_{N_R}$.
The left, middle, and right panels show $\Delta_{D,\,ij}^{N_R}$, $\Delta_{D^*,\,ij}^{N_R}$ and $\Delta_{\Lambda_c,\,ij}^{N_R}$, respectively. 
Different colors correspond to the different operator combinations, as indicated in the legend of the left panel.
In the left panel, the $V'_RS'_R$ curve coincides with $V'_RS'_L$, and $V'_RV'_R$ lies close to $S'S'$. 
In the middle and right panels, the scalar curves are degenerate. 
}
\label{fig:delta_X_ij}
\end{figure}
%%%%%%%%%%%%%%%%%%%%%%%%%%%%%%%%%%%%%%%

We introduce the parameters $\Delta_{H_c}^{N_R}$ and $\Delta_{H_c,\,ij}^{N_R}$, defined as
\begin{align}
 \Delta_{H_c}^{N_R} = \sum_{ij} \Delta_{H_c,\,ij}^{N_R}\,C_i\,C_j^{*} \equiv \frac{R^\prime_{H_c}}{R_{H_c}^{\rm SM}}-1.~~~
 (i,j = V_R^\prime, S_L^\prime, S_R^\prime, T^\prime)
 \label{eq:deltaNR}
\end{align}
These parameters vanish in the SM limit, $R_{H_c}^\prime = R_{H_c}^{\rm SM}$.
In the presence of NP contributions, although they do not vanish in the small-velocity limit, $\delta^{N_R}$ and $\delta_{ij}^{N_R}$ can be written as
\begin{align}
 \delta^{N_R} &= \Delta_{\Lambda_c}^{N_R} - \frac{1}{4} \Delta_{D}^{N_R} -  \frac{3}{4} \Delta_{D^*}^{N_R}, \notag \\
 \delta_{ij}^{N_R} &= \Delta_{\Lambda_c,\,ij}^{N_R} - \frac{1}{4} \Delta_{D,\,ij}^{N_R} -  \frac{3}{4} \Delta_{D^*,\,ij}^{N_R}.
 \label{eq:deltaNRij}
\end{align}
Therefore, the violation of the sum rule can be evaluated as a weighted sum of these parameters. 
Figure~\ref{fig:delta_X_ij} shows $\Delta_{D,\,ij}^{N_R}$ (left), $\Delta_{D^*,\,ij}^{N_R}$ (middle), $\Delta_{\Lambda_c,\,ij}^{N_R}$ (right) as functions of $m_{N_R}$ for various operators.
Here and in the following, $(ij) = (S^\prime S^\prime)$ denotes both $(ij) = (S_R^\prime S_R^\prime)$ and $(S_L^\prime S_L^\prime)$, since $\Delta^{N_R}_{S_R^\prime S_R^\prime}=\Delta^{N_R}_{S_L^\prime S_L^\prime}$ is satisfied.
All contributions except for the interference terms are positive, since the $N_R$ emission amplitudes do not interfere with the SM contribution.
It is also observed that $\Delta_{H_c,\,ij}^{N_R}$ are suppressed as the sterile neutrino becomes heavier due to the phase space suppression.
In the presence of the tensor operator, $(ij)=(T^\prime T^\prime)$ and $(V_R^\prime T^\prime)$, we find that $\Delta^{N_R}_{D^*,\,ij}$ and $\Delta^{N_R}_{\Lambda_c,\,ij}$ can be as large as about $15$ and $-5$.
However, the remaining contributions are at most of order $2$.

%%%%%%%%%%%%%%%%%%%%%%%%%%%%%%%%%%%%%%%
\begin{figure}[t]
\begin{center}
\includegraphics[width=0.445\linewidth]{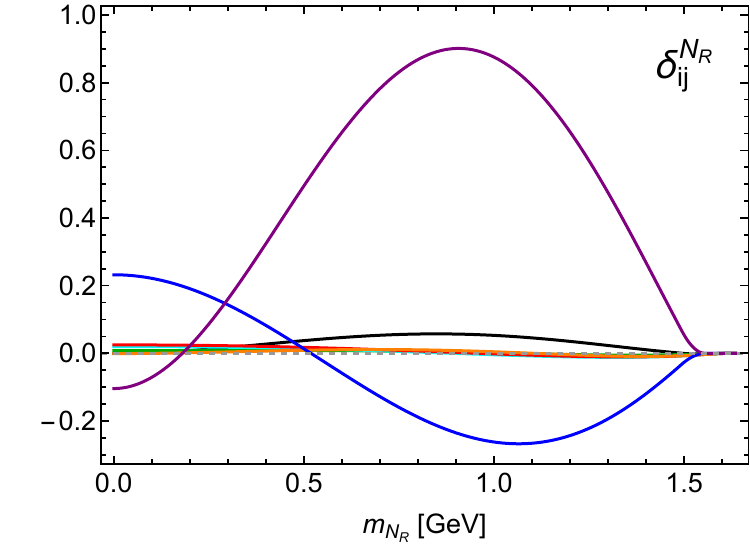}~
\includegraphics[width=0.45\linewidth]{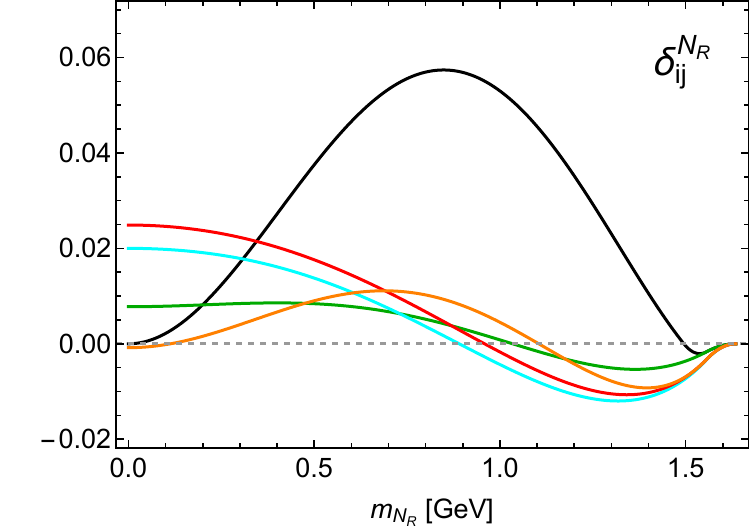}\\
\end{center}
\vspace{-.15cm}
\caption{
The sum rule violation coefficients $\delta^{N_R}_{ij}$ as functions of $m_{N_R}$.
Different colors correspond to different operator combinations, as indicated in the legend of the left panel of Fig.~\ref{fig:delta_X_ij}. 
Several curves overlap in the left panel, and the right panel shows a magnified view.}
\label{fig:delta_ij}
\end{figure}
%%%%%%%%%%%%%%%%%%%%%%%%%%%%%%%%%%%%%%%

In Fig.~\ref{fig:delta_ij}, we show the deviation $\delta^{N_R}_{ij}$ for each pair $(ij)$.
The definition of each color is given in Fig.~\ref{fig:delta_X_ij}.
As anticipated from Fig.~\ref{fig:delta_X_ij}, the deviation can be enhanced for the $V_R^\prime T^\prime$ (blue) and $T^\prime T^\prime$ (purple) contributions.
The right panel is a magnified view of the left one, with these contributions omitted.
We observe that only the $V^\prime_R T^\prime$ interference term can give a sizably negative contribution to $\delta^{N_R}$.
We also find that the $T^\prime T^\prime$ and $V_R^\prime V_R^\prime$ contributions peak around 0.8\,GeV, while the $V_R^\prime T^\prime$ contribution peaks around 1.1\,GeV.

%%%%%%%%%%%%%%%%%%%%%%%%%%%%%%%%%%%%%%%
\begin{figure}[t]
\begin{center}
\includegraphics[width=0.44\linewidth]{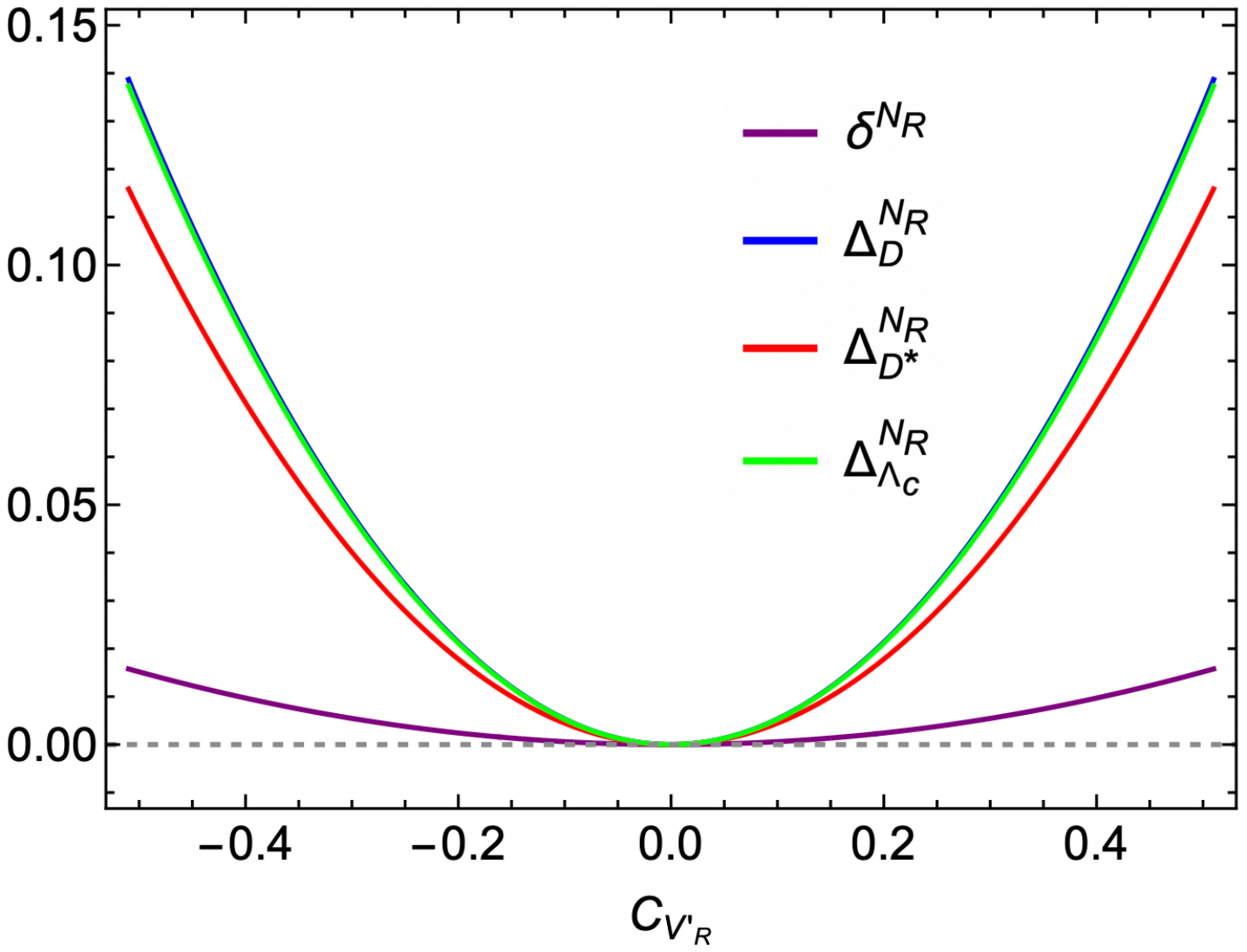}~
\includegraphics[width=0.46\linewidth]{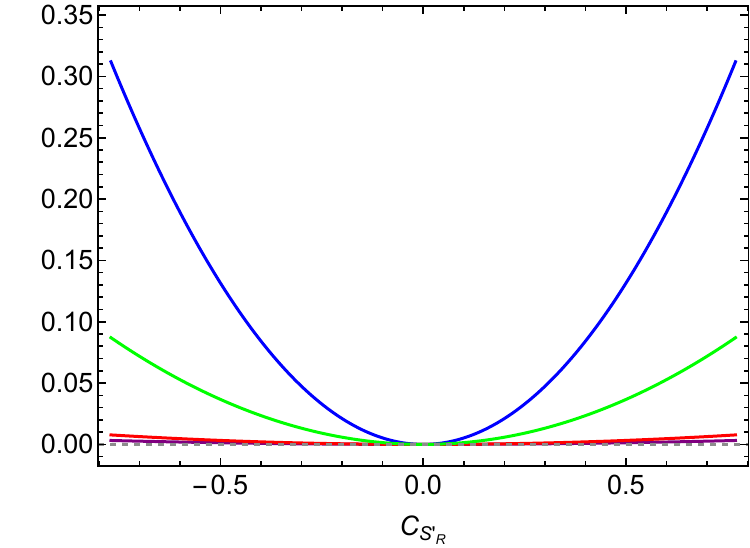}\\ \vskip .12in
\includegraphics[width=0.45\linewidth]{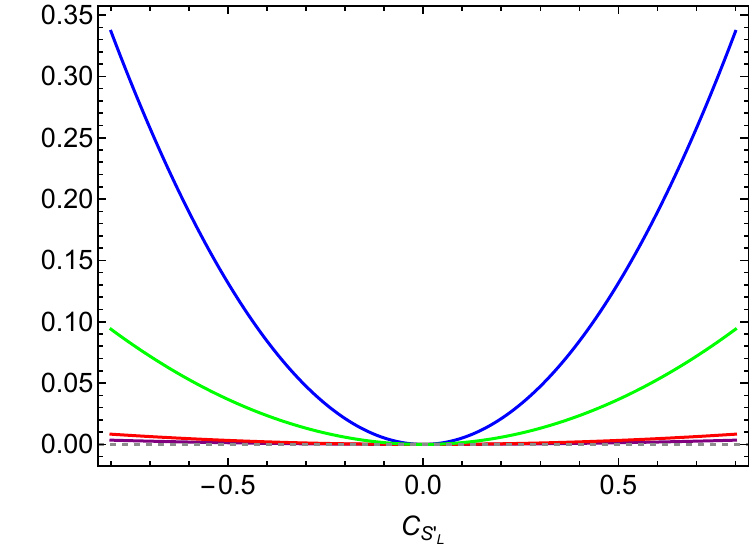}~
\includegraphics[width=0.445\linewidth]{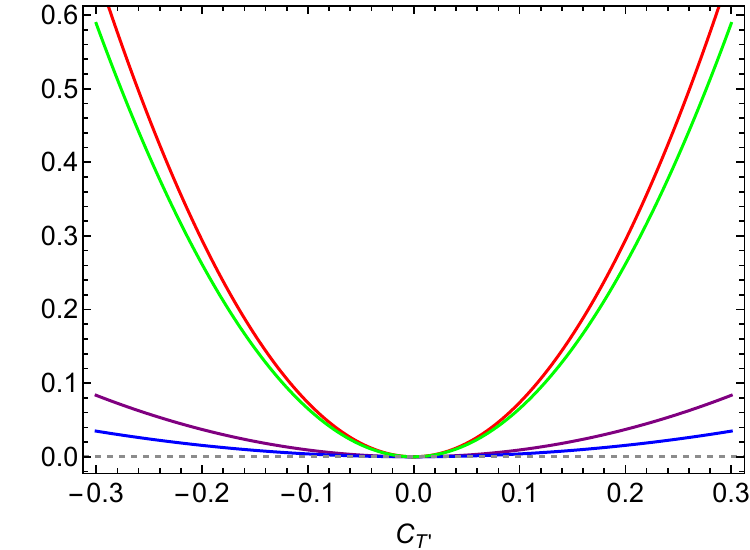}\\
\end{center}
\vspace{-.15cm}
\caption{
The sum rule violation $\delta^{N_R}$ and parameters $\Delta_{H_c}^{N_R}$ as functions of the WCs in single operator scenarios.
The sterile neutrino mass is fixed to $m_{N_R}=0.8$\,GeV.
The boundaries of the horizontal axis correspond to the maximally allowed values of the WCs under the relaxed LHC constraint.
}
\label{fig:delta_ij_NP}
\end{figure}
%%%%%%%%%%%%%%%%%%%%%%%%%%%%%%%%%%%%%%%

Figure \ref{fig:delta_ij_NP} shows the results obtained in the single-operator scenario, in which only one of the Wilson coefficients $C_i$ is switched on while the others are set to zero.
The range of the horizontal axis is determined using the relaxed LHC constraint.\footnote{
The WCs are more tightly constrained under the stringent LHC bound, as shown in Table~\ref{tab:LHC_bound}.
}
Under these constraints, the violation of the sum rule $\delta^{N_R}$ (purple) is at most of order $0.1$, since the allowed values of $C_i$ are smaller than unity (see Table~\ref{tab:LHC_bound}).
Comparing with $\Delta_{H_c}^{N_R}$, we observe a significant cancellation among the contributions from $H_c=D$ (blue), $D^*$ (red), and $\Lambda_c$ (green), resulting in the net quantity $\delta^{N_R}$ (purple).
Here the sterile neutrino mass is fixed to $0.8$\,GeV, for which we find that $\delta^{N_R}$ remains positive.
The tensor contribution can reach $\delta^{N_R} \simeq 0.08$, whereas the other operators yield $\delta^{N_R} \lesssim 0.02$.
As shown in Fig.~\ref{fig:delta_ij}, $\delta^{N_R}$ can become negative for $m_{N_R} > 1$\,GeV.
However, after multiplying by the corresponding Wilson coefficients, the magnitude of $\delta^{N_R}$ is reduced to below $0.01$.

As mentioned in the Introduction, the current experimental values $R_{H_c}$ imply $\delta\simeq-0.41\pm0.24$ when interpreted under the standard assumption of a massless neutrino emission. 
This motivates examining whether sterile neutrino emission can drive $\delta^{N_R}$ negative and thereby affect the apparent tension.\footnote{
In general, the signal acceptance for sterile neutrino emission differs from that for SM neutrino emission, and it is not obvious to what extent the sterile neutrino channel contributes to the measurements of $R_{H_c}$.
The following discussion should therefore be regarded as illustrative.
} 
As shown in Fig.~\ref{fig:delta_ij}, $(ij) = (V^\prime_R T^\prime)$ can yield a sizable negative contribution.
However, this interference term is accompanied by additional contributions from $(ij) = (V^\prime_R V^\prime_R)$ and $(T^\prime T^\prime)$.
To quantitatively assess how large the deviation can become, we consider a scenario in which both $C_{V_R^\prime}$ and $C_{T^\prime}$ are non-vanishing, while $C_{S_L^\prime}=C_{S_R^\prime}=0$.
For the LHC constraints, the interference between the vector and tensor operators has only a small impact on the high-$p_T$ tails of the signal events. 
We therefore impose the following constraint on the Wilson coefficients,
\begin{align}
    \left(\frac{C_{V_R^\prime}}{C_{V_R^\prime}^{\rm max}}\right)^2+\left(\frac{C_{T^\prime}}{C_{T^\prime}^{\rm max}}\right)^2<1,
    \label{eq:LHCmax}
\end{align}
where $(C_{V_R^\prime}^{\rm max},\,C_{T^\prime}^{\rm max})=(0.51,\,0.30)$ under the relaxed LHC constraint, and $(C_{V_R^\prime}^{\rm max},\,C_{T^\prime}^{\rm max})=(0.33,\,0.17)$ under the stringent LHC bound.
Under this constraint, we determine $(C_{V_R^\prime},\,C_{T^\prime})$ that minimizes $\delta^{N_R}$.

%%%%%%%%%%%%%%%%%%%%%%%%%%%%%%%%%%%%%%%
\begin{figure}[t]
\begin{center}
\includegraphics[width=0.43\linewidth]{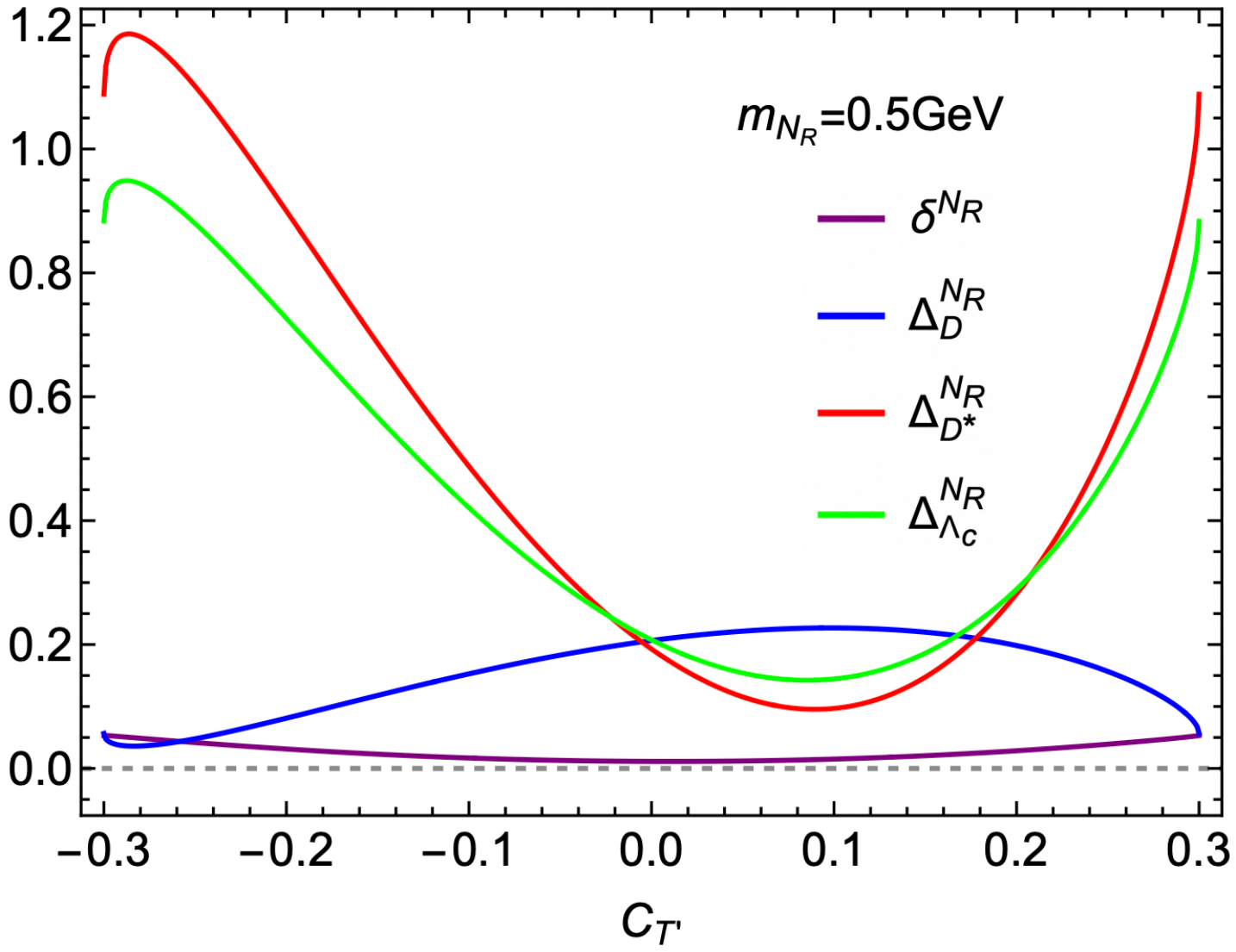}~
\includegraphics[width=0.455\linewidth]{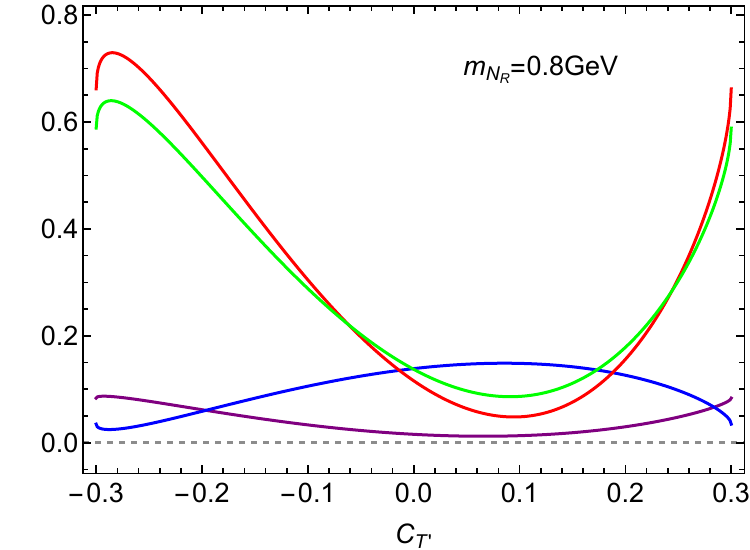}\\\vskip .12in
\includegraphics[width=0.445\linewidth]{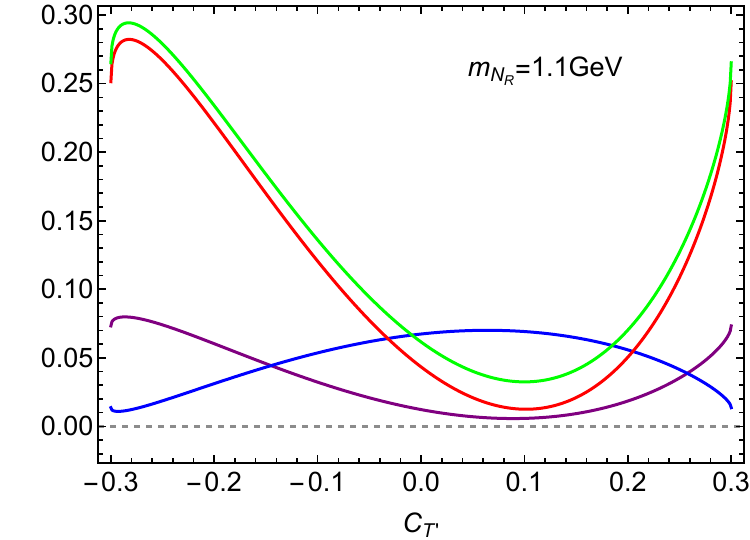}~
\includegraphics[width=0.45\linewidth]{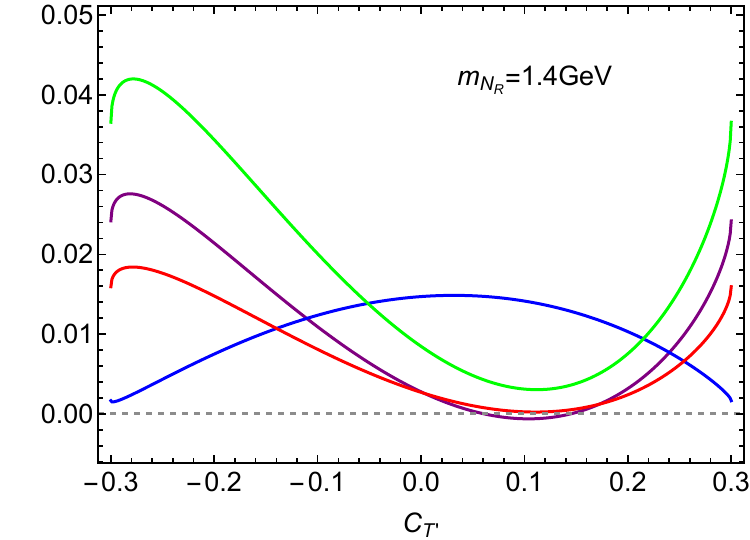}\\
\end{center}
\vspace{-.15cm}
\caption{
The sum rule violation $\delta^{N_R}$ and parameters $\Delta_{H_c}^{N_R}$ as functions of $C_{T^\prime}$ in the scenario in which both $C_{V_R^\prime}$ and $C_{T^\prime}$ are non-vanishing, while $C_{S_L^\prime}=C_{S_R^\prime}=0$.
The boundaries of the horizontal axis correspond to the maximally allowed values under the relaxed LHC constraint.
}
\label{fig:delta_ij_VRT}
\end{figure}
%%%%%%%%%%%%%%%%%%%%%%%%%%%%%%%%%%%%%%%

In Fig.~\ref{fig:delta_ij_VRT}, we show the deviation $\delta^{N_R}$ for $m_{N_R}=0.5$, $0.8$, $1.1$, and $1.4\,{\rm GeV}$.
The range of the horizontal axis is determined using the relaxed LHC constraint.
For simplicity, we keep $C_{V_R^\prime}$ positive and allow $C_{T^\prime}$ to take either sign.
We observe that even if $\Delta^{N_R}_{D^*}$ and $\Delta^{N_R}_{\Lambda_c}$ are of $\mathcal{O}(1)$, the total $\delta^{N_R}$ remains at the level of $\sim 0.1$ after the cancellation among the contributions $\Delta^{N_R}_{H_c}$.\footnote{
We note that for $\Delta_{D^*}^{N_R}\simeq 1$, $R^\prime_{D^*}/R^{\rm SM}_{D^*}$ becomes approximately 2, which is inconsistent with the measurement if the acceptance is assumed to be the same as that of the SM signal. }
The deviation drops sharply around $C_i \simeq -0.3$, where the interference contribution becomes suppressed.
Although $\delta^{N_R}_{V_R^\prime T^\prime}$ takes the largest negative value around $m_{N_R}=1.1$\,GeV in Fig.~\ref{fig:delta_ij}, the total $\delta^{N_R}$ remains positive because of the positive contributions from $(ij) = (V_R^\prime V_R^\prime)$ and $(T^\prime T^\prime)$.
We find that $\delta^{N_R}$ can become negative around $m_{N_R}=1.4$\,GeV.
However, its magnitude is very small, since $\Delta^{N_R}_{H_c}$ are strongly suppressed by phase space and therefore negligible compared with the current experimental uncertainty. 
In conclusion, we find that the sum rule is robust against the presence of a sterile neutrino.

As the sterile neutrino becomes heavier, some of the decay channels may become kinematically forbidden.
The decays $B\to D^*\tau\bar N_R$ and $\Lambda_b\to \Lambda_c\tau\bar N_R$ are forbidden for $m_{N_R} \gtrsim 1.49$ and $1.56$\,GeV, respectively, while $B\to D\tau\bar N_R$ remains kinematically allowed.
Although the sum rule is then explicitly violated, the phase space suppression is significant, and the NP contributions themselves become almost invisible even in the measurements of $R_{H_c}$.

\begin{figure}[t]
\begin{center}
\includegraphics[width=0.306\linewidth]{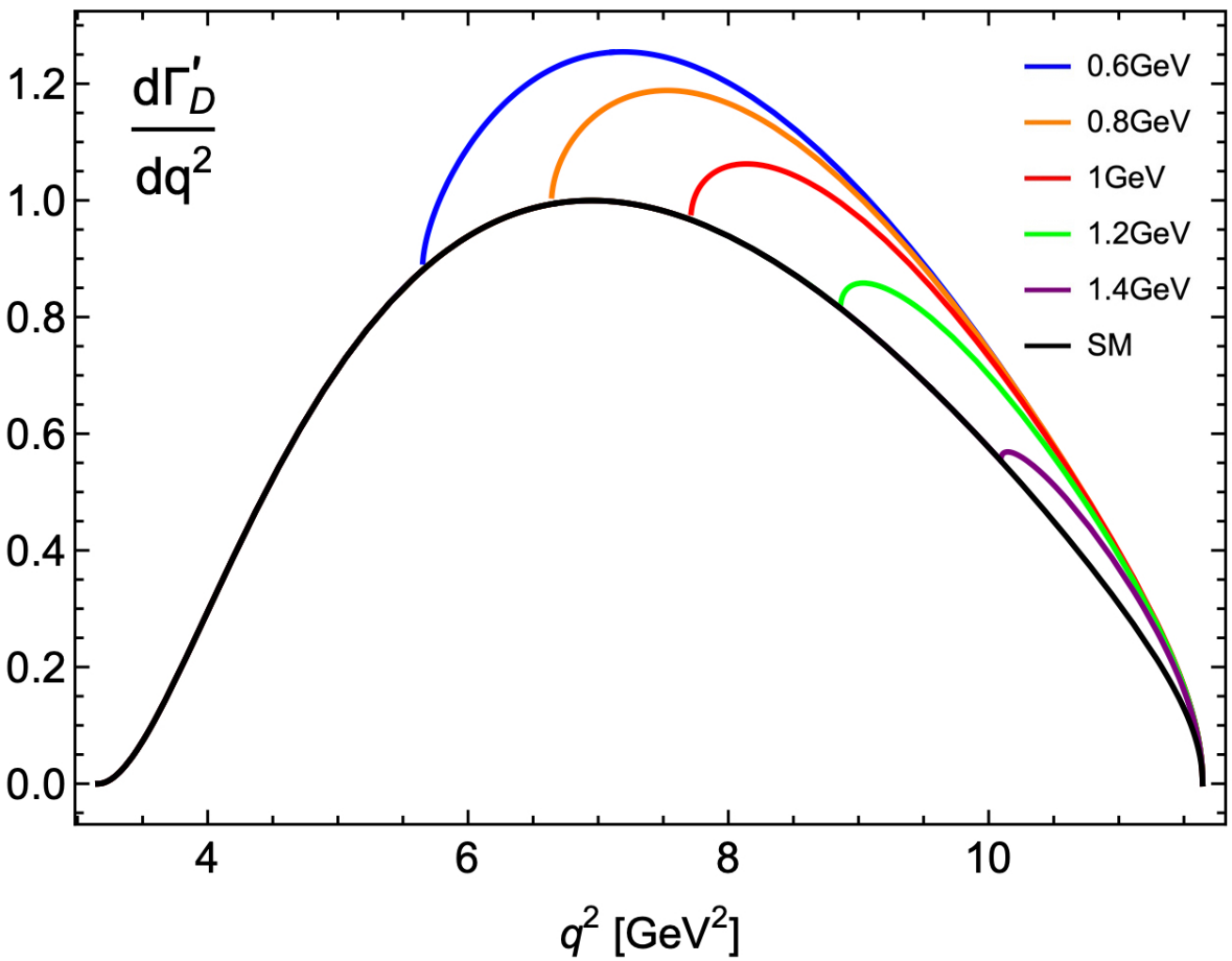}~
\includegraphics[width=0.318\linewidth]{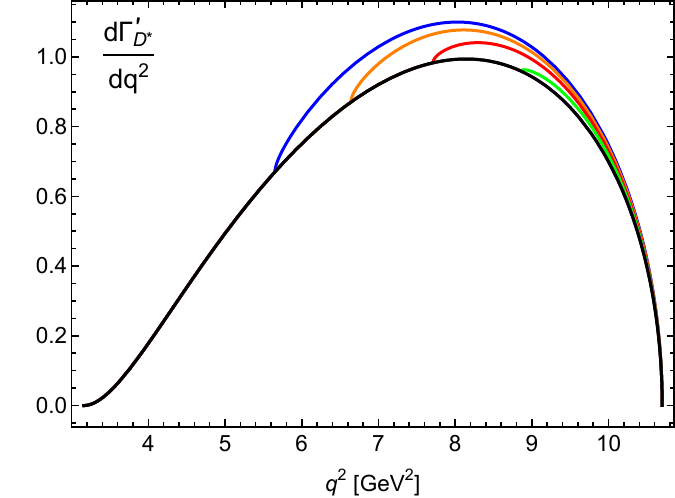}~
\includegraphics[width=0.318\linewidth]{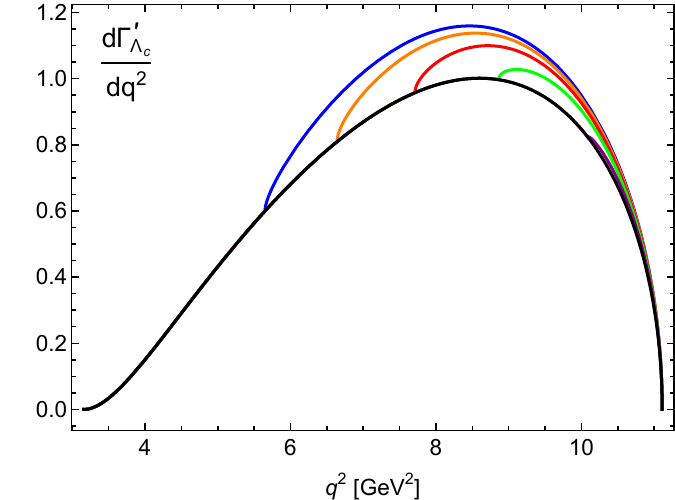}
\end{center}
\vspace{-.15cm}
\caption{
Normalized differential distributions for $B\to D\tau \bar X$, $B\to D^* \tau\bar X$ and $\Lambda_b\to\Lambda_c\tau \bar X$ (from left to right), where $X=\nu$ or $N_R$. 
The normalization is chosen such that the peak of the SM distribution is unity. 
}
\label{fig:diff}
\end{figure}

Although we have confirmed that the contributions from the sterile neutrino are almost negligible for the sum rule, the particle may still be probed through detailed studies of differential decay rates.
Such an additional massive sterile neutrino modifies the kinematic distributions of the visible final states.
As an illustration, we continue with the scenario in which both $C_{V_R^\prime}$ and $C_{T^\prime}$ are non-vanishing, while $C_{S_L^\prime}=C_{S_R^\prime}=0$.
We set $C_{T^\prime}=0.1$, which corresponds to the minimum of $\delta^{N_R}$ while keeping $\Delta^{N_R}_{H_c}$ at the level of about $0.2$ in all cases.
Figure \ref{fig:diff} shows the normalized differential distributions for the decays $B\to D\tau \bar X$, $B\to D^*\tau \bar X$, and $\Lambda_b\to\Lambda_c\tau \bar X$ where $X=\nu$ or $N_R$. 
Here, the normalization is chosen such that the maximum of the SM distribution is unity.
In the presence of a sterile neutrino, the differential decay rates start to deviate from the SM prediction (black) above the kinematic threshold $q^2_{\rm min}=(m_\tau+m_{N_R})^2$. 
The effect of the sterile neutrino decreases with increasing mass due to stronger phase space suppression.
To test such a scenario, measurements of differential distributions are required.

%%%%%%%%%%%%%%%%%%%%%%%%%%%%%%%%%%%%%%%
%%%%%%%%%%%%%%%%%%%%%%%%%%%%%%%%%%%%%%%

So far, we have focused on the deviation $\delta^{N_R}$.
However, we do not attempt a quantitative recast of the experimental measurements of $R_{H_c}$ in the presence of sterile neutrino emission, as no experimental results are directly applicable to this scenario. 
The signal regions are typically defined in kinematical regions where the backgrounds are efficiently suppressed.
Since the kinematics of the final-state particles, including a massive sterile neutrino, differs from that in the case of a massless (left-handed) neutrino, the experimental values of $R_{H_c}$, which are reported under the assumption that a massless neutrino is emitted, cannot be applied to Eq.~\eqref{eq:RSRHQw2} straightforwardly.
Nevertheless, a naive fit to the current experimental values of $R_{D}$ and $R_{D^{*}}$, ignoring the acceptance differences, suggests that for $m_{N_R} \lesssim 0.8$\,GeV, sterile neutrino contributions can improve the fit to the current data, but do not resolve the mismatch among $R_D$, $R_{D^*}$, and $R_{\Lambda_c}$ in the sum rule. 
For larger masses, the best-fit values of the WCs exceed the relaxed LHC bound, and imposing that constraint leads to a deterioration of the fit quality.
A dedicated analysis accounting for the modified acceptance is left for future work.

Before closing this section, we comment on the constraints taking into account the SU(2)$_L$ invariance.
Since the left-handed fermions belong to SU(2)$_L$ doublets in the SM,
flavor-changing neutral-current (FCNC) processes such as $b\to s \nu \bar N_R$ can be induced.
To see this, it is convenient to express the effective operators in the SU(2)$_L$ invariant form as
\begin{align}
 &O_{V_R^{\prime\prime}} = (\overline{c} \gamma^\mu P_R b)(\overline{\tau} \gamma_\mu P_R N_R), \,\,\, 
 O_{S_R^{\prime\prime}} = \epsilon(\overline{Q_2}  P_R b)(\overline{L_3} P_R N_R), \label{eq:operator2}\notag \\
 &O_{S_L^{\prime\prime}} = (\overline{c} P_L Q_3)(\overline{L_3} P_R N_R), \,\,\, 
 O_{T^{\prime\prime}} = \epsilon(\overline{Q_2}  \sigma^{\mu\nu}P_R b)(\overline{L_3} \sigma_{\mu\nu} P_R N_R),
\end{align}
where $\epsilon$ denotes the anti-symmetric tensor.
We work in the down-quark mass basis, $Q^T= (V^\dagger u_L,\,d_L)$, where $V$ is the CKM matrix.
For instance, $Q_3^T= (V^*_{ib}\,u_i,\,b)$ and $\bar{Q}_2=(V_{i s}\bar{u}_i,\,\bar{s})$.
The operators relevant for the neutral-current transitions are then given by
\begin{align}
 &O_{V_R^\prime}^{n} = {\rm null\,process}, \,\,\,\,\,\,
 O_{S_R^\prime}^{n} = (\overline{s}  P_R b)(\overline{\nu_\tau} P_R N_R), \label{eq:operator3}\notag \\
 &O_{S_L^\prime}^{n} = V_{u_i b}^*(\overline{c} P_L u_i)(\overline{\nu_\tau} P_R N_R), \,\,\,\,\,\,
 O_{T^\prime}^{n} = (\overline{s}  \sigma^{\mu\nu}P_R b)(\overline{\nu_\tau} \sigma_{\mu\nu} P_R N_R).
\end{align}
The $V_R^\prime$ operator does not involve left-handed quark fields and therefore does not induce FCNC processes at tree level.
In contrast, down-type neutral currents are generated by $S_R^\prime$ and $T^\prime$.
The corresponding Wilson coefficients are subject to stringent upper bounds of $\mathcal{O}(10^{-2})$ \cite{Bernlochner:2024xiz,Felkl:2023ayn,Kamenik:2009kc,Kamenik:2011vy,Rosauro-Alcaraz:2024mvx,Robinson:2018gza,Buras:2014fpa,Belle-II:2023esi}.
Since the available phase space is larger in $b\to s$ transitions, the effect of the $N_R$ mass is expected to be mild.
Therefore, the realistic size of the sum rule violation is likely to be smaller, resulting in a more robust sum rule.
On the other hand, the current bounds involving $c\to d \nu \bar N_R$ are weaker.

%%%%%%%%%%%%%%%%%%%%%%%%%%%%%%%%%%%%%
\section{Conclusion and discussion}
\label{sec:conclusion}
%%%%%%%%%%%%%%%%%%%%%%%%%%%%%%%%%%%%%

In this paper, we examined whether a massive sterile neutrino can invalidate the $b \to c$ semileptonic sum rule.
We derived the sum rule for total decay rates in the small velocity limit and evaluated its violation for realistic hadrons.
Although the sterile neutrino modifies the decay rates and can close the relevant decay channels at different thresholds, the resulting violation remains negligible compared with the present experimental uncertainties.
This result shows that the sterile neutrino loophole is effectively closed within the current setup.

Beyond the integrated rates, a massive sterile neutrino leaves a characteristic imprint on the $q^2$ distribution of the differential decay rate, opening an independent avenue for its detection.
A quantitative assessment of this signal requires a dedicated evaluation of the experimental acceptance, which depends on $m_{N_R}$ and differs from that of the SM neutrino mode, and is left for future work in close collaboration with experimentalists.
It would further be interesting to extend the analysis to doubly differential decay rates involving angular distributions (cf.~Ref.~\cite{Endo:2025cvu}), providing additional handles for disentangling sterile neutrino contributions.

%%%%%%%%%%%%%%%%%%%%%%%%%%%%%%%%%%%%%%%%%%%%%%%%%%%%%%%
\section*{Acknowledgements}
%%%%%%%%%%%%%%%%%%%%%%%%%%%%%%%%%%%%%%%%%%%%%%%%%%%%%%%
We appreciate the members of Theory-Experiment Assembly (TEA), participants of its workshop at Gamagori, Ryoutaro Watanabe, Florian Kretz and Hiroyasu Yonaha for stimulating discussion.
This work is supported by JSPS KAKENHI Grant Numbers 22K21347 [M.E. and S.I.], 24K07025 [S.M.], 24K22879 [S.I.], 24K23939 [S.I.], 25K17385 [S.I.], and by the Deutsche Forschungsgemeinschaft (DFG, German Research Foundation) under grant 396021762 - TRR 257 for the Collaborative
Research Center \textit{Particle Physics Phenomenology after the Higgs Discovery (P3H)} [T.K.]. 
The work is also supported by the Toyoaki scholarship foundation [S.I.].
We also appreciate KEK-KMI joint appointment program [M.E. and S.I.], which boosted this project. 
%%%%%%%%%%%%%%%%%%%%%%%%%%%%%%%%%%%%%%%%%%%%%%%%%%%%%%%% 
\appendix

%%%%%%%%%%%%%%%%%%%%%%%%%%%%%%%%%%%%%
\section{Hadronic and leptonic amplitudes}
\label{sec:amp}
%%%%%%%%%%%%%%%%%%%%%%%%%%%%%%%%%%%%%
In this appendix, we describe the hadronic and leptonic amplitudes. 
The analysis of the $B \to D^{(*)}$ transitions is based on Refs.~\cite{Bernlochner:2017jka,Iguro:2020cpg}.
In the HQET, the hadron matrix elements are parametrized as  
\begin{align}
 \langle D | \bar c \gamma^\mu b | B \rangle 
 & = \sqrt{m_B m_D} \big[ h_+ (v+v')^\mu + h_- (v-v')^\mu \big], 
  \\[0.5em]
 \langle D |\bar c b| B \rangle
 & = \sqrt{m_B m_D} (w+1) h_S, 
 \\[0.5em]
 \langle D |\bar c \gamma^\mu \gamma_5 b| B \rangle
 & = \langle D |\bar c \gamma_5 b| B \rangle = 0, 
  \\[0.5em]
 \langle D |\bar c \sigma^{\mu\nu} b| B \rangle
 & = -i \sqrt{m_B m_D}\, h_T \big( v^\mu v^{\prime\nu} - v^{\prime\mu} v^\nu  \big), 
  \\[0.5em]
 \langle D^* | \bar c \gamma^\mu b | B \rangle
 & = i \sqrt{m_B m_{D^*}} h_V \varepsilon^{\mu\nu\rho\sigma} \epsilon^*_\nu v'_\rho v_\sigma, 
  \\[0.5em]
 \langle D^* | \bar c \gamma^\mu \gamma_5 b | B \rangle 
 & = \sqrt{m_B m_{D^*}} \big[ h_{A_1} (w+1) \epsilon^{*\mu} - (\epsilon^* \cdot v) \left( h_{A_2} v^\mu + h_{A_3} v^{\prime\mu} \right) \big], 
  \\[0.5em]
  \langle D^* | \bar c \gamma_5 b | B \rangle 
 & = -\sqrt{m_B m_{D^*}} (\epsilon^* \cdot v) h_P, 
  \\[0.5em]
 \langle D^* |\bar c b| B \rangle
 & = 0, 
  \\[0.5em]
 \langle D^* | \bar c \sigma^{\mu\nu} b | B \rangle 
 & = -\sqrt{m_B m_{D^*}} \varepsilon^{\mu\nu\rho\sigma} 
 \big[ h_{T_1} \epsilon^{*}_\rho (v+v')_\sigma + h_{T_2} \epsilon^{*}_\rho (v-v')_\sigma 
 \notag \\ 
 & \qquad\qquad\qquad\qquad~
 + h_{T_3} (\epsilon^* \cdot v) (v+v')_\rho (v-v')_\sigma \big],
\end{align} 
with $v^\mu = p_B^\mu /m_B$ and $v^{\prime\mu} = p_{D^{(*)}}^\mu /m_{D^{(*)}}$.
For $q^2=(p_B-p_{D^{(*)}})^2$, $w =v \cdot v' = (m_B^2+m_{D^{(*)}}^2-q^2)/(2m_Bm_{D^{(*)}})$ varies in the range of $1 \leq w \leq w_{D^{(*)}}^{\rm max}$ with $w_{D^{(*)}}^{\rm max} = (m_B^2+m_{D^{(*)}}^2-(m_\tau+m_{N_R})^2)/(2m_Bm_{D^{(*)}})$.
The form factors $h_X$ are functions of $w$ and expressed in the heavy quark limit by the leading order IW function $\xi(w)$ as~\cite{Isgur:1989vq}
\begin{align}
 \label{eq:DFFinHQL}
 & h_+ = h_V = h_{A_1} = h_{A_3} = h_S = h_P = h_T = h_{T_1} = \xi(w), \\
 & h_- = h_{A_2} = h_{T_2} = h_{T_3} = 0.
\end{align}
The IW function satisfies $\xi(1)=1$.
Departing from the heavy quark limit, the form factors include corrections.
Defining $\hat h_X(w) \equiv h_X(w)/\xi(w)$, they are generally expanded as
\begin{align}
 \hat h_X = \hat h_{X,0} + \frac{\alpha_s}{\pi} \delta \hat h_{X,\alpha_s} + \frac{\bar \Lambda}{2m_b} \delta \hat h_{X,m_b} + \frac{\bar \Lambda}{2m_c} \delta \hat h_{X,m_c} + \left(\frac{\bar \Lambda}{2m_c}\right)^2 \delta \hat h_{X,m_c^2}, 
 \label{eq:deltaFF}
\end{align} 
where $\hat h_{X,0} = 1$ for $X = +,V,A_1,A_3,S,P,T,T_1$ and $0$ for $X = -,A_2,T_2,T_3$, denoting the leading order contributions.
Also, $\bar \Lambda$ is a QCD scale.
The corrections $\delta \hat h_X$ are taken into account at $\mathcal{O}(\alpha_s, \bar \Lambda/m_{b,c}, \bar \Lambda^2/m_{c}^2)$ by following Refs.~\cite{Bernlochner:2017jka,Iguro:2020cpg}.

Similarly to the $B \to D^{(*)}$ transitions, the HQET form factors for the $\Lambda_b \to \Lambda_c$ transitions are given by~\cite{Bernlochner:2018bfn}
\begin{align}
 \langle \Lambda_c| \bar c\gamma_\mu b |\Lambda_b\rangle 
 &= \bar u(p',s') \big[ f_1 \gamma_\mu + f_2 v_\mu + f_3 v'_\mu \big] u(p,s), 
 \\[0.5em]
 \langle \Lambda_c| \bar c\gamma_\mu\gamma_5 b |\Lambda_b\rangle 
 &= \bar u(p',s') \big[ g_1 \gamma_\mu + g_2 v_\mu + g_3 v'_\mu \big] \gamma_5\, u(p,s), 
 \\[0.5em]
 \langle \Lambda_c| \bar c\, b |\Lambda_b\rangle 
 &= h'_S\, \bar u(p',s')\, u(p,s), 
 \\[0.5em]
 \langle \Lambda_c| \bar c \gamma_5 b |\Lambda_b\rangle 
 &= h'_P\, \bar u(p',s')\, \gamma_5\, u(p,s), 
 \\[0.5em]
 \langle \Lambda_c| \bar c\, \sigma_{\mu\nu}\, b |\Lambda_b\rangle 
 &= \bar u(p',s') \big[ h_1\, \sigma_{\mu\nu}
 + i\, h_2 (v_\mu \gamma_\nu - v_\nu \gamma_\mu)
 + i\, h_3 (v'_\mu \gamma_\nu - v'_\nu \gamma_\mu)
 \notag \\ 
 & \qquad\qquad~
 + i\, h_4 (v_\mu v'_\nu - v_\nu v'_\mu) \big] u(p,s),
\end{align}
where $u(p,s)$ are spinors with momentum $p$ and spin $s$.
Also, $v = p/m_{\Lambda_b}$, $v' = p'/m_{\Lambda_c}$, $w = v \cdot v' = (m_{\Lambda_b}^2 + m_{\Lambda_c}^2 - q^2)/(2m_{\Lambda_b} m_{\Lambda_c})$, and $w_{\Lambda_c}^{\rm max} = (m_{\Lambda_b}^2+m_{\Lambda_c}^2-(m_\tau+m_{N_R})^2)/(2m_{\Lambda_b}m_{\Lambda_c})$ are introduced.
The form factors $f_i$, $g_i$, and $h^{(\prime)}_i$ are functions of $w$ and expressed in the heavy quark limit as~\cite{Isgur:1990pm}
\begin{align}
 \label{eq:LambdaFFinHQL}
 & f_1 = g_1 = h'_S = h'_P = h_1 = \zeta(w), \\
 & f_2 = f_3 = g_2 = g_3 = h_2 = h_3 = h_4 = 0,
\end{align}
where $\zeta(w)$ is the IW function for ground state baryons, satisfying $\zeta(1)=1$.
Once departing from the heavy quark limit, the $\Lambda_b \to \Lambda_c$ form factors include corrections similar to Eq.~\eqref{eq:deltaFF}. 
They are taken into account at $\mathcal{O}(\alpha_s, \bar \Lambda/m_{b,c}, \alpha_s\bar \Lambda/m_{b,c}, \bar \Lambda^2/m_{c}^2)$ by following Refs.~\cite{Bernlochner:2018kxh,Bernlochner:2018bfn}. 

In the following, we summarize the hadronic factors of the helicity amplitudes in the HQET. 
For the $B \to D$ transitions, there are four amplitudes~\cite{Iguro:2020cpg,Sakaki:2013bfa}:
\begin{align}
 H_{V,0}^D &= m_B \sqrt{\frac{r_D(w^2-1)}{\hat q_D^2}} \Big[ (1+r_D)h_+ -(1-r_D)h_- \Big], \\
 H_{V,t}^D &= m_B \sqrt{\frac{r_D}{\hat q_D^2}} \Big[ (1-r_D)(w+1)h_+ -(1+r_D)(w-1)h_- \Big], \\
 H_{S}^D &= m_B \sqrt{r_D} (w+1) h_S, \\
 H_{T}^D &= - m_B \sqrt{r_D(w^2-1)}\, h_T.
\end{align}
For the $B \to D^*$ transitions, they are given by~\cite{Iguro:2020cpg, Sakaki:2013bfa}
\begin{align}
 H_{V,\pm}^{D^*} 
 &= m_B \sqrt{r_{D^*}} \Big[ (w+1) h_{A_1} \mp \sqrt{w^2-1} h_{V} \Big], \\ 
 H_{V,0}^{D^*} 
 &= m_B \sqrt{\frac{r_{D^*}}{\hat q_{D^*}^2}} (w+1) 
 \Big[ (r_{D^*}-w) h_{A_1} + (w-1) (r_{D^*} h_{A_2} + h_{A_3}) \Big], \\
 H_{V,t}^{D^*} 
 &= -m_B \sqrt{\frac{r_{D^*}(w^2-1)}{\hat q_{D^*}^2}}
 \Big[ (w+1) h_{A_1} + (r_{D^*}w-1) h_{A_2} + (r_{D^*}-w) h_{A_3} \Big], \\
 H_{P}^{D^*} 
 &= - m_B \sqrt{r_{D^*}(w^2-1)} h_P, \\ 
 H_{T,\pm}^{D^*} 
 &= \pm m_B \sqrt{\frac{r_{D^*}}{\hat q_{D^*}^2}} \left[ 1-r_{D^*} (w \mp \sqrt{w^2-1}) \right]
 \left[ h_{T_1} + h_{T_2} + \left( w \pm \sqrt{w^2-1} \right) (h_{T_1} - h_{T_2}) \right], \\
 H_{T,0}^{D^*} 
 &= -m_B \sqrt{r_{D^*}} \Big[ (w+1) h_{T_1} +(w-1) h_{T_2} +2(w^2-1) h_{T_3} \Big].
\end{align}
For $\Lambda_b \to \Lambda_c$, the hadronic factors are shown as~\cite{Datta:2017aue, Bernlochner:2018bfn}
\begin{align}
 H_{V,0\pm}^{\Lambda_c} 
 &= m_{\Lambda_b} \sqrt{\frac{2r_\Lambda}{\hat q_\Lambda^2}} 
 \Big\{
 \sqrt{w-1} \Big[(1+r_\Lambda)f_1 + (w+1)(f_2 r_\Lambda + f_3)\Big] \\ 
 & \qquad\qquad\quad
 \mp \sqrt{w+1} \Big[(1-r_\Lambda)g_1 - (w-1)(g_2 r_\Lambda + g_3)\Big]
 \Big\},  \notag \\
 H_{V,1\pm}^{\Lambda_c} 
 &= -2 \,m_{\Lambda_b} \sqrt{r_\Lambda} \Big[ \sqrt{w-1} f_1 \mp \sqrt{w+1} g_1 \Big], \\
 H_{V,t\pm}^{\Lambda_c} 
 &= m_{\Lambda_b} \sqrt{\frac{2r_\Lambda}{\hat q_\Lambda^2}} \Big\{
 \sqrt{w+1} \Big[(1-r_\Lambda)f_1 + f_2(1-w r_\Lambda) + f_3(w-r_\Lambda)\Big]
\\ 
 & \qquad\qquad\quad
 \mp \sqrt{w-1} \Big[(1+r_\Lambda)g_1 - g_2(1-w r_\Lambda) - g_3(w-r_\Lambda)\Big] \Big\}, 
  \notag \\
 H_{S}^{\Lambda_c} 
 &= m_{\Lambda_b} \sqrt{2r_\Lambda(w+1)} h'_S, \\
 H_{P}^{\Lambda_c} 
 &= m_{\Lambda_b} \sqrt{2r_\Lambda(w-1)} h'_P, \\
 H_{T,0\pm}^{\Lambda_c} 
 &= m_{\Lambda_b} \sqrt{2r_\Lambda} \Big\{ \sqrt{w-1} \Big[h_1-h_2+h_3-(w+1)h_4\Big] \pm \sqrt{w+1}h_1 \Big\},\\
 H_{T,1\pm}^{\Lambda_c}
 &= -2 m_{\Lambda_b} \sqrt{\frac{r_\Lambda}{\hat q_\Lambda^2}} 
 \Big\{ 
 \sqrt{w-1} \Big[ (1+r_\Lambda) h_1 - (1-wr_\Lambda) h_2 - (w-r_\Lambda) h_3 \Big] 
 \notag \\ 
 & \qquad\qquad\qquad
 \pm \sqrt{w+1} \Big[ (1-r_\Lambda) h_1 - (w-1) (h_2 r_\Lambda + h_3) \Big] 
 \Big\},
\end{align}
where $\hat q_{H_c}^2 = q^2/m_{H_b}^2 = 1 - 2 r_{H_c} w + r^2_{H_c}$ and $r_{H_c} = m_{H_c}/m_{H_b}$.

The leptonic amplitudes involving a massive sterile neutrino are given by 
\begin{align}
    L_{S'}(\lambda_\tau,\lambda_N) &= \langle \tau(p_\tau,\lambda_\tau) \bar{N}_R(p_N,\lambda_N) | \bar{\tau} P_R N_R | 0 \rangle, \\
    L_{V'}(\lambda_\tau,\lambda_N,\lambda_W) &= \epsilon_\mu(q,\lambda_W) \langle \tau(p_\tau,\lambda_\tau) \bar{N}_R(p_N,\lambda_N) | \bar{\tau} \gamma^\mu P_R N_R | 0 \rangle, \\
    L_{T'}(\lambda_\tau,\lambda_N,\lambda_W,\lambda_{W'}) &= \epsilon_\mu(q,\lambda_W) \epsilon_\nu(q,\lambda_{W'}) \langle \tau(p_\tau,\lambda_\tau) \bar{N}_R(p_N,\lambda_N) | \bar{\tau} \sigma^{\mu\nu}P_R N_R | 0 \rangle.
\end{align}
For a given helicity configuration, the leptonic amplitudes for the scalar operator read
\begin{align}
    L_{S'}(\pm, \pm) = \mp \frac{1}{\sqrt{2}} \sqrt{q^2-m_\tau^2-m_{N_R}^2 \mp \sqrt{Q^l_+ Q^l_-}}.
\end{align}
For the vector operator, the leptonic amplitudes read
\begin{align}
    L_{V'}(\pm,\pm,+)& = -L_{V'}(\pm,\pm,-) \\
    &= - \frac{\sin\theta}{2 \sqrt{2} \sqrt{q^2}} \sqrt{(q^2)^2- \bigg(m_\tau^2 - m_{N_R}^2 \pm \sqrt{Q^l_+ Q^l_-} \bigg)^2}, \notag \\
    L_{V'}(\pm,\pm,t)& = \mp \frac{1}{2 \sqrt{q^2}} \sqrt{(q^2)^2- \bigg(m_\tau^2 - m_{N_R}^2 \pm \sqrt{Q^l_+ Q^l_-} \bigg)^2},  \\
    L_{V'}(\pm,\pm,0)& = - \frac{\cos\theta}{2 \sqrt{q^2}} \sqrt{(q^2)^2- \bigg(m_\tau^2 - m_{N_R}^2 \pm \sqrt{Q^l_+ Q^l_-} \bigg)^2}, \\
    L_{V'}(\pm,\mp,0)& = \mp \frac{\sin\theta}{\sqrt{2}} \sqrt{q^2- m_\tau^2 - m_{N_R}^2 \pm \sqrt{Q^l_+ Q^l_-} }, \\
    L_{V'}(\pm,\mp,\mp)& = - \cos^2\bigg( \frac{\theta}{2} \bigg) \sqrt{q^2- m_\tau^2 - m_{N_R}^2 \pm \sqrt{Q^l_+ Q^l_-} }, \\
    L_{V'}(\pm,\mp,\pm)& = - \sin^2\bigg( \frac{\theta}{2} \bigg) \sqrt{q^2- m_\tau^2 - m_{N_R}^2 \pm \sqrt{Q^l_+ Q^l_-} },
\end{align}
and the tensor one reads
\begin{align}
     L_{T'}(\pm,\pm,\pm,\mp)  & = -L_{T'}(\pm,\pm,\mp,\pm) = \pm L_{T'}(\pm,\pm,0,t) = \mp L_{T'}(\pm,\pm,t,0)  \\
     &= \pm \frac{i}{\sqrt{2}} \cos\theta \sqrt{q^2 - m_\tau^2 - m_{N_R}^2 \mp \sqrt{Q^l_+ Q^l_-}},\notag \\
     L_{T'}(\pm,\mp,\pm,\mp)& = - L_{T'}(\pm,\mp,\mp,\pm) =  \pm L_{T'}(\pm,\mp,0,t) = \mp L_{T'}(\pm,\mp,t,0)  \\
     &= i \frac{\sin\theta}{2 \sqrt{q^2}} \sqrt{(q^2)^2 - \bigg( m_\tau^2 - m_{N_R}^2 \mp \sqrt{Q^l_+ Q^l_-} \bigg)^2}, \notag \\
     L_{T'}(\pm,\pm,\pm,0)  & = L_{T'}(\pm,\pm,\mp,0) = -L_{T'}(\pm,\pm,0,\pm) = - L_{T'}(\pm,\pm,0,\mp) \\
     &= \pm  L_{T'}(\pm,\pm,\pm,t) = \mp L_{T'}(\pm,\pm,\mp,t) = \mp L_{T'}(\pm,\pm,t,\pm)  \notag\\
     &= \pm L_{T'}(\pm,\pm,t,\mp) \notag\\
     &= \frac{i}{2} \sin\theta \sqrt{q^2-m_\tau^2 -m_{N_R}^2 \mp \sqrt{Q^l_+ Q^l_-}},\notag\\
     L_{T'}(\pm,\mp,\pm,0) &= -L_{T'}(\pm,\mp,0,\pm) = \pm L_{T'}(\pm,\mp,\pm,t) = \mp L_{T'}(\pm,\mp,t,\pm) \\
     &= \pm \frac{i}{\sqrt{2} \sqrt{q^2}} \sin^2\bigg(\frac{\theta}{2}\bigg) \sqrt{(q^2)^2 - \bigg( m_\tau^2 -m_{N_R}^2 \mp \sqrt{Q^l_+ Q^l_-} \bigg)^2},\notag\\
     L_{T'}(\pm,\mp,\mp,0) &= -L_{T'}(\pm,\mp,0,\mp) = \mp L_{T'}(\pm,\mp,\mp,t)  = \pm L_{T'}(\pm,\mp,t,\mp) \\
     &= \mp \frac{i}{\sqrt{2}\sqrt{q^2}} \cos^2\bigg(\frac{\theta}{2}\bigg) \sqrt{(q^2)^2 - \bigg( m_\tau^2 -m_{N_R}^2 \mp \sqrt{Q^l_+ Q^l_-} \bigg)^2}. \notag
\end{align}
The leptonic angle $\theta$ is defined as the angle between the charged lepton and the $B$ meson momentum in the $W$-boson rest frame.

%%%%%%%%%%%%%%%%%%%%%%%%%%%%%%%%%%%%%%%%%%%%%%%%%%
\bibliographystyle{utphys28mod}
\bibliography{ref}
\end{document}